\documentclass[preprint,12pt]{elsarticle}

\usepackage{amssymb}
\usepackage{amsmath} 

\journal{Journal of Neuroscience Methods}

\begin{document}

\begin{frontmatter}

\title{Deep Learning Models for Atypical Serotonergic Cells Recognition}

\author[inst1,inst2]{Daniele Corradetti}

\affiliation[inst1]{organization={Grupo de Fisica Matematica, Instituto Superior Tecnico},
            addressline={Av. Rovisco Pais}, 
            city={Lisboa},
            postcode={1049-001}, 
            country={Portugal}}

\affiliation[inst2]{organization={Departamento de Matematica, Universidade do Algarve},
            addressline={Campus de Gambelas}, 
            city={Faro},
            postcode={8005-139}, 
            state={Faro},
            country={Portugal}}
\author{Alessandro Bernardi}

\author[inst3]{Renato Corradetti}
\affiliation[inst3]{organization={Department of Neuroscience, Psychology, Drug Research and Child Health (NEUROFARBA), University of Florence},
            addressline={Viale G. Pieraccini 6}, 
            city={Firenze},
            postcode={50139}, 
            state={Toscana},
            country={Italy}}

\begin{abstract}

\emph{Background}: The serotonergic system modulates brain processes via functionally distinct
subpopulations of neurons with heterogeneous properties, including their electrophysiological
activity. In extracellular recordings, serotonergic neurons to be investigated for their functional properties are commonly identified on the basis of “typical” features of their activity, i.e. slow regular firing and relatively long duration of action potentials. Thus, due to the lack of equally robust criteria for discriminating serotonergic neurons with “atypical” features from non-serotonergic cells, the physiological relevance of the diversity of serotonergic neuron activities results largely understudied.

\emph{New Methods}: We propose deep learning models capable of discriminating typical and atypical
serotonergic neurons from non-serotonergic cells with high accuracy. The research utilized
electrophysiological \emph{in vitro} recordings from serotonergic neurons identified by the expression of fluorescent proteins specific to the serotonergic system and non-serotonergic cells. These recordings formed the basis of the training, validation, and testing data for the deep learning models. The study employed convolutional neural networks (CNNs), known for their efficiency in pattern recognition, to classify neurons based on the specific characteristics of their action potentials.

\emph{Results}: The models were trained on a dataset comprising 27,108 original action potential samples, alongside an extensive set of 12 million synthetic action potential samples, designed to mitigate the risk of overfitting the background noise in the recordings, a potential source of bias. Results show that the models achieved high accuracy and were further validated on ”non-homogeneous” data, i.e., data unknown to the model and collected on different days from those used for the training of the model, to confirm their robustness and reliability in real-world experimental conditions.

\emph{Comparison with existing methods}: Conventional methods for identifying serotonergic neurons
allow recognition of serotonergic neurons defined as typical. Our model based on the analysis of the sole action potential reliably recognizes over 94\% of serotonergic neurons including those with atypical features of spike and activity. 

\emph{Conclusions}: The model is ready for use in experiments conducted with the here described
recording parameters. We release the codes and procedures allowing to adapt the model to different acquisition parameters or for identification of other classes of spontaneously active neurons.
\end{abstract}


\end{frontmatter}


\section{Introduction}

Activity of serotonergic neurons is known to regulate a wealth of
autonomic and higher functions in mammals (Steinbusch et al., 2021;
Faulkner and Deakin, 2014; Pilowsky, 2014; Lesch et al., 2012; Monti,
2011). Present knowledge of the physiological and pharmacological
properties of serotonergic neurons is mostly based on electrophysiological
recordings of neuronal activity from raphe nuclei of laboratory animals
both \emph{in vivo} and \emph{in vitro}. However, most of recordings
have been performed on neurons whose serotonergic identity was based
on criteria that were empirically developed in the years to restrict
the investigations to recordings from neurons that displayed very
typical activity. For serotonergic neurons, the accepted criteria
require the concomitant regularity of firing, broad action potential
and, when pharmacological assays were allowed by the experimental
design, sensitivity to serotonin1A receptor agonists that typically
produce reversible slowing or cessation of neuron firing. When recordings
are conducted in slices under microscopy guidance, the large size
of serotonergic neuron soma could be used as an additional criterion.
Adhering to these strict criteria for serotonergic neuron identification,
however, results in a selection bias that has limited the studies
to the \textquotedblleft typical\textquotedblright{} neurons which
might underrepresent the variety of serotonergic neuron population.
The implication of serotonin neuron activity in behavioural tasks in mice has been studied using selective optogenetic activation and recording from raphe neurons \emph{in vivo} (Liu et al., 2014) as well as with one-photon calcium imaging (Paquelet et al., 2022). Interestingly, the latter study revealed anatomically defined subpopulations of DRN serotonin neurons with different activity and projecting to either reward-related or anxiety-related brain areas.
This confirmed previous evidence for the existence of subpopulations of serotonergic
neurons with distinctive neurochemical and pharmacological properties
as well as firing patterns, emerged in the course of the past 40 years
of dedicated research (e.g. Calizo et al., 2011; Paquelet et al., 2022;
see also in Gaspar et al., 2012; Andrade and Haj-Dahmane, 2013; Commons, 2020). For instance, using \emph{in vitro} recordings from dorsal raphe nucleus the possibility
that serotonergic neurons display also irregular firing or peculiar
rhythmic fluctuations in firing activity has been described since
early recordings both \emph{in vivo} and \emph{in vitro} (Mosko and
Jacobs (1974, 1976) and more recently confirmed with recordings of
serotonergic neurons from transgenic mice selectively expressing fluorescent
proteins in serotonergic neurons (Mlinar et al, 2016). Thus, the principal
drawback of the intra-experiment recognition of 5-HT neurons is that
serotonergic neurons displaying atypical activity or spikes narrower
than expected are discarded and their pharmacological and physiological
characteristics remain elusive. In addition, in the course of our
research on genetically fluorescent serotonergic neurons (Montalbano
et al., 2015; Mlinar 2016) we also noticed the existence of non-serotonergic
(non-fluorescence labelled) neurons with regular activity and relatively
broad spikes whose duration often overlaps that of action potentials recorded
in serotonergic neurons. Thus, in \textquotedblleft real life\textquotedblright{}
experimental conditions the activity characteristics of a non-neglectable
number of serotonergic and non-serotonergic neurons could overlap
and adherence to the above-mentioned strict criteria for identification
of typical serotonergic neurons has the advantage to ensure a reasonable
homogeneity of the population under study, in spite of the selection
bias introduced. On the other hand, the characteristics of what we
define ``atypical'' serotonergic neurons remain understudied. 

In the present work we have taken advantage of the recordings present
in our internal database and obtained from transgenic mice selectively
expressing fluorescent proteins in serotonergic neurons to develop
deep-learning based models for recognition of serotonergic and non-serotonergic
neurons with relatively high accuracy and that can be implemented
in the recording programs to quickly help the experimenter in the
decision of continuing the recording or to change the experimental
design, should an atypical serotonergic or non-serotonergic neuron
be identified. 

\section{Material and Methods}

\subsection{Source database}

To train, test and validate our deep-learning based models we used
the original recordings from our internal database built in the occasion
of our studies in which we described the firing characteristics of
genetically identified dorsal raphe serotonergic neurons in brain
slices. Serotonergic and non-serotonergic neurons were thus identified
on the basis of a parameter independent from their electrophysiological
features, i.e., on serotonergic system-specific fluorescent protein
expression (serotonergic) or lack of expression (non-serotonergic).
In our original articles (Mlinar et al., 2016; Montalbano et al.,
2015) we detailed the procedure to obtain the three transgenic mouse
lines with serotonergic system-specific fluorescent protein expression
used in the present work: Tph2::SCFP; Pet1-Cre::Rosa26.YFP ; Pet1-Cre::CAG.eGFP. 

\subsection{Loose-seal cell-attached recordings}

Detailed description of the electrophysiological methods and of the
measures for improving reliability of loose-seal cell-attached recordings
has been previously published (Montalbano et al., 2015; Mlinar et
al., 2016). In brief, mice (4-28 weeks of age) were anesthetized with
isofluorane and decapitated. The brains were rapidly removed and dissected
in ice-cold gassed (95\% O2 and 5\% CO2) ACSF composed of: 124 mM
NaCl, 2.75 mM KCl, 1.25 mM NaH2PO4, 1.3 mM MgCl2, 2 mM CaCl2, 26 mM
NaHCO3, 11 mM D-glucose. The brainstem was sliced coronally into 200
\textmu m thick slices with a vibratome (DSK, T1000, Dosaka, Japan).
Slices were allowed to recover for at least 1 h at room temperature
and then were individually transferred to a submersion type recording
chamber and continuously superfused at a flow rate of 2 ml min-1 with
oxygenated ACSF warmed to 37°C by a feedback-controlled in-line heater
(TC-324B / SF-28, Warner Instruments, Hamden, CT). Slices were allowed
to equilibrate for 10-20 min before the beginning of the recording.
To reproduce in brain slices noradrenergic drive that facilitates
serotonergic neuron firing during wakefulness (Baraban and Aghajanian,
1980; Levine and Jacobs, 1992), ACSF was supplemented with the natural
agonist noradrenaline (30 $\mu$M) or with the $\alpha$1
adrenergic receptor agonist phenylephrine (10 $\mu$M; Vandermaelen
and Aghajanian, 1983). Neurons within DRN were visualized by infrared
Dodt gradient contrast video microscopy, using a 40X water-immersion
objective (N-Achroplan, numerical aperture 0.75, Zeiss, Göttingen,
Germany) and a digital CCD camera (ORCA-ER C4742-80-12AG; Hamamatsu,
Hamamatsu City, Japan) mounted on an upright microscope (Axio Examiner
Z1; Zeiss) controlled by Axiovision software (Zeiss). Loose-seal cell-attached
recordings were made from fluorescent protein-expressing or not expressing
neurons, visually identified by using Zeiss FilterSet 46 (eGFP and
YFP, excitation BP 500/20, emission BP 535/30) or Zeiss FilterSet
47 (CFP, excitation BP 436/20, emission BP 480/40). Fluorescence was
excited using a Zeiss HXP 120 lamp. Patch electrodes (3-6 M$\Omega$)
were pulled from thick-walled borosilicate capillaries (1.50 mm outer
diameter, 0.86 mm inner diameter; Corning) on a P-97 Brown-Flaming
puller (Sutter Instruments, Novato, CA) and filled with solution containing
(in mM): 125 NaCl, 10 HEPES, 2.75 KCl, 2 CaCl2 and 1.3 MgCl2, pH 7.4
with NaOH. After positioning the pipette, development of loose-seal
was monitored by using a voltage-clamp protocol with holding potential
of 0 mV and test pulse of 1 mV / 100 ms, repeated every second. Weak
positive pressure was released and gentle suction was slowly applied
until detected spikes increased to 50 - 100 pA peak-to-peak amplitude.
In some experiments this procedure was repeated during recording to
increase signal to noise ratio. Corresponding seal resistance was
in 10 to 20 M$\Omega$ range. Recordings were made using an Axopatch
200B amplifier (Molecular Devices, Sunnyvale, CA) controlled by Clampex
9.2 software (Molecular Devices). Signals were low-pass filtered with
a cut-off frequency of 5 kHz (Bessel) and digitized with sampling
rate of 40 kHz (Digidata 1322A, Molecular Devices). After the recording,
images of recorded neuron were acquired to document the expression
of the fluorescent marker in the recorded neuron. 

\subsection{Offline Analysis of recordings}

Detection of spikes was performed using event detection routine of
Clampfit 9.2 software. Spike duration (width) was determined from
the shape of averaged action potential by measuring the interval between the
spike upstroke and the downstroke (or second downstroke, whenever
present) hereby named UDI (Upstroke-Downstroke Interval) for convenience
(see Fig. 6; see also Fig. 3 in Mlinar et al., 2016).

\section{A Deep Learning Model}

Recognizing serotonergic cells is a binary classification problem,
i.e., serotonergic vs. non-serotonergic cells, for which deep learning
(DL) algorithms and, more specifically, the use of convolutional neural
networks (CNN) have yielded excellent results. Notably, CNN are inspired
by the organization of the animal visual system, particularly the
human brain, and excel at tasks like image feature extraction, which
is fundamental for recognition purposes (Liu, 2018). They employ mechanisms
such as feedforward inhibition to alleviate issues like gradient vanishing,
enhancing their effectiveness in complex pattern recognition tasks
(Liu et al, 2019). With these considerations in mind, we have chosen
to use a CNN architecture even in the apparently unconventional context
of numerical pattern recognition, i.e., the recorded signal of a neuronal
cell. The inspiring idea behind this choice is to leverage the ability
of CNNs to amplify numerical patterns that occur at different scales,
in this case within time intervals that are orders of magnitude smaller
than the entire examined signal. In fact, this is a characteristic
typical of neuronal spikes, where the maximum peak impulse can occur
within a scale of 1 ms, while the firing period, i.e., the time interval
between two consecutive spikes, can be two orders of magnitude greater.

\subsection{Preliminary approaches and definition of appropriate parameters for developing the model}

Starting from the assumption that two factors are typically relevant
in recognizing serotonergic cells, namely the specific shape of the
action potential together with its repetitiveness and firing frequency, we initially
decided to consider time segments of 7 seconds as training data for
the neural network. This ensured an adequate number of action potentials to evaluate
their consistency and periodicity. After several attempts in this
direction, however, we realized that the importance of the cell's
action potential shape was so predominant that the information obtained from
analyzing the firing periodicity alone was not sufficient to compensate
for the accuracy gained by focusing on the individual action potential. 

Our first preliminary analysis was done on 108 serotonergic cells
and 45 non-serotonergic cells. Every action potential for the training consisted
in the recording of 7 ms taken from 2 ms before the detection threshold
to 5 ms after. While the final accuracy of the resulting models was
fairly high, ranging from 94.3\% to 99.3\%, further analysis on non-homogeneous
data, i.e. data from neurons whose identity was kept unknown to the model and were collected on experimental days different from those used for the training and evaluation of the models, showed a much lower accuracy, which was a strong sign of the overfitting. Further
investigation allowed to identify an important source of overfitting
in the background noise of the recordings which, having a specific
signature, the model learned to incorporate in the recognition of
the neuron types. Thus, models trained with action potentials embedded in 7 ms
time-segments learned how to classify the spikes on the basis of the
background noise instead of the peculiar shape of the event. 

Therefore, we decided to reduce the impact of the background noise
present in the samples by limiting the time-window of action potential analysis
to 4 ms. This solution worked well, since we had a comparable accuracy
of the metrics on non-homogeneous data. 

 Another very efficient solution for expanding the training data, beside splitting the samples in different segments, was  given by the generation of a synthetic data set for which we develop a very specific procedure (see section 3.2) that combines smoothed action potentials signals along with real noise masks. To this purpose, we produced 12M synthetic action potentials from a pool of 600 different noise backgrounds, thus reducing the impact
that such noise could have in the training.  The training on synthetic data led to an improvement on all accuracy types on non-homogeneous data (e.g. from binary accuracy 0.9125 to 0.9375, from AUC 0.8976 to  0.9255 and from F1-Score 0.8679 to 0.9056, see Fig. \ref{fig:Confusion-matrix-of} for more details). Besides the specific improvement in model performance, it is important to note the utility of the synthetic model in monitoring sources of overfitting arising from noise signatures in the recordings. More specifically, the difference in accuracy on non-homogeneous data between the biological model and the synthetic model provides a rough estimate of the overfitting in the biological model resulting from noise signatures. This is highly significant when determining how additional experiments with different noise signatures could improve the model. 

\begin{figure}
\centering{}\includegraphics[scale=0.75]{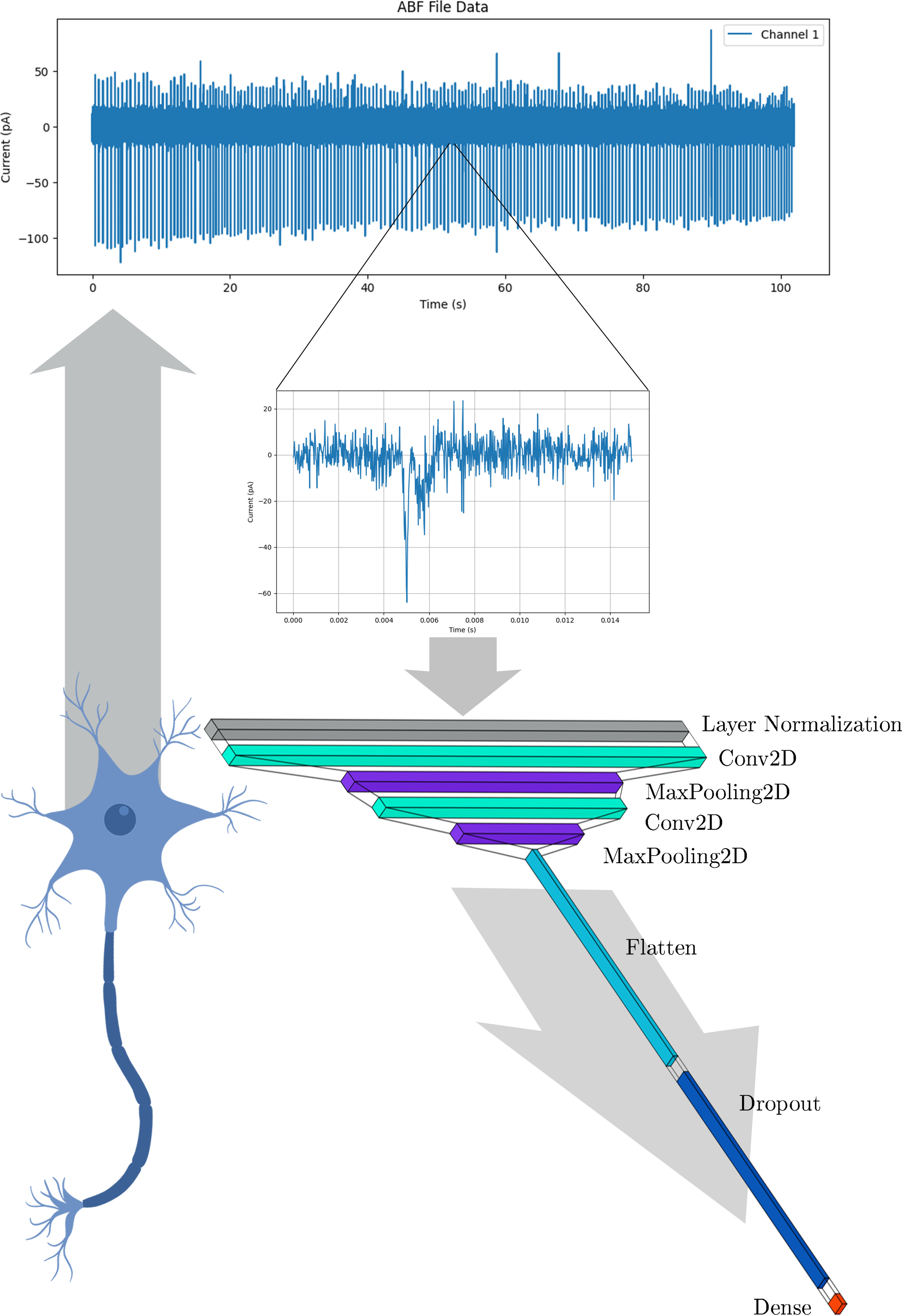}\caption{Summary of the various steps used to implement the model from recorded
signals: from neuronal cell the signal is sampled at 40 kHz and recorded
as .abf file, then the all events are selected and sent to 10 neural
networks with the above architecture for classification (only difference
between the architectures is the value of the 2D convolutional kernel
with ranges between 20 to 30).}
\end{figure} 
\subsection{Data used for originating and validating the final model}

\paragraph{Original Training Data}

The original data for the training, validation and testing of the
models consisted in 43,327 action potential samples extracted from 108 serotonergic
cells and 45 non-serotonergic cells.  Since the two classes were unbalanced (29,773 serotonergic and 13,554 non-serotonergic) we undersampled the serotonergic class, to obtain a more balanced dataset for training.  Therefore, the training set data consisted in 13,554 action potentials from serotonergic cells, and 13,554 action potentials from non-serotonergic cells. In all cases, the triggering threshold of the event was -50 pA and the spike was then sampled 1 ms before the triggering threshold until 3 ms after (see Figs. 1,2). Since the sampling rate of the original
recordings was 40 kHz, every action potential sample consists of 160 values.
All the samples were then randomly subdivided into 18,975 for training,
4,066 for validation and 4,067 for testing. 
\begin{figure}
\begin{centering}
\includegraphics[scale=0.7]{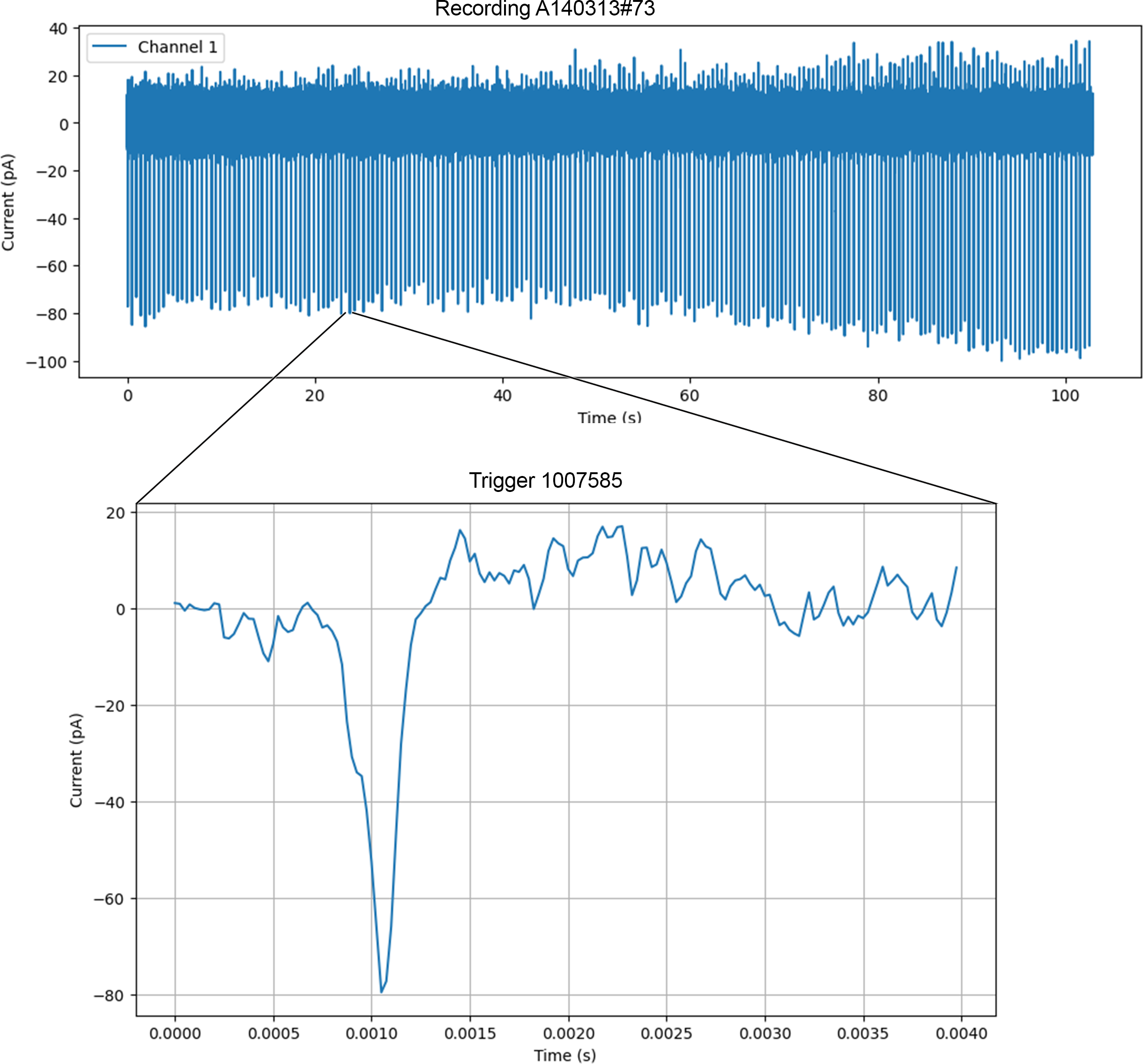}
\par\end{centering}
\centering{}\caption{Example on how single events were isolated and selected. The image
depicts the recording of the serotonergic cell A140313\#073 and the
4 ms event of triggered at point 1007585, i.e. at 25.189 sec.}
\end{figure}

\paragraph{Non-homogeneous Data}

The non-homogeneous data consisted in 24,616 samples extracted from a new set of 55 serotonergic cells (18,595 action potentials) and 27 non-serotonergic cells (6,021 action potentials) collected in experimental days not used to obtain the training data, thus with different signal noise. These data were never part of the training set, nor validation, nor testing set during the training. Furthermore, the identity of the neurons from which these data were obtained was unknown to the model. Non-homogeneous data were therefore used as an additional, independent, test for the already trained model to assess its robustness when cells have a noise signal never encountered by the model.

\paragraph{Synthetic Data}

\begin{figure}
\includegraphics[scale=0.8]{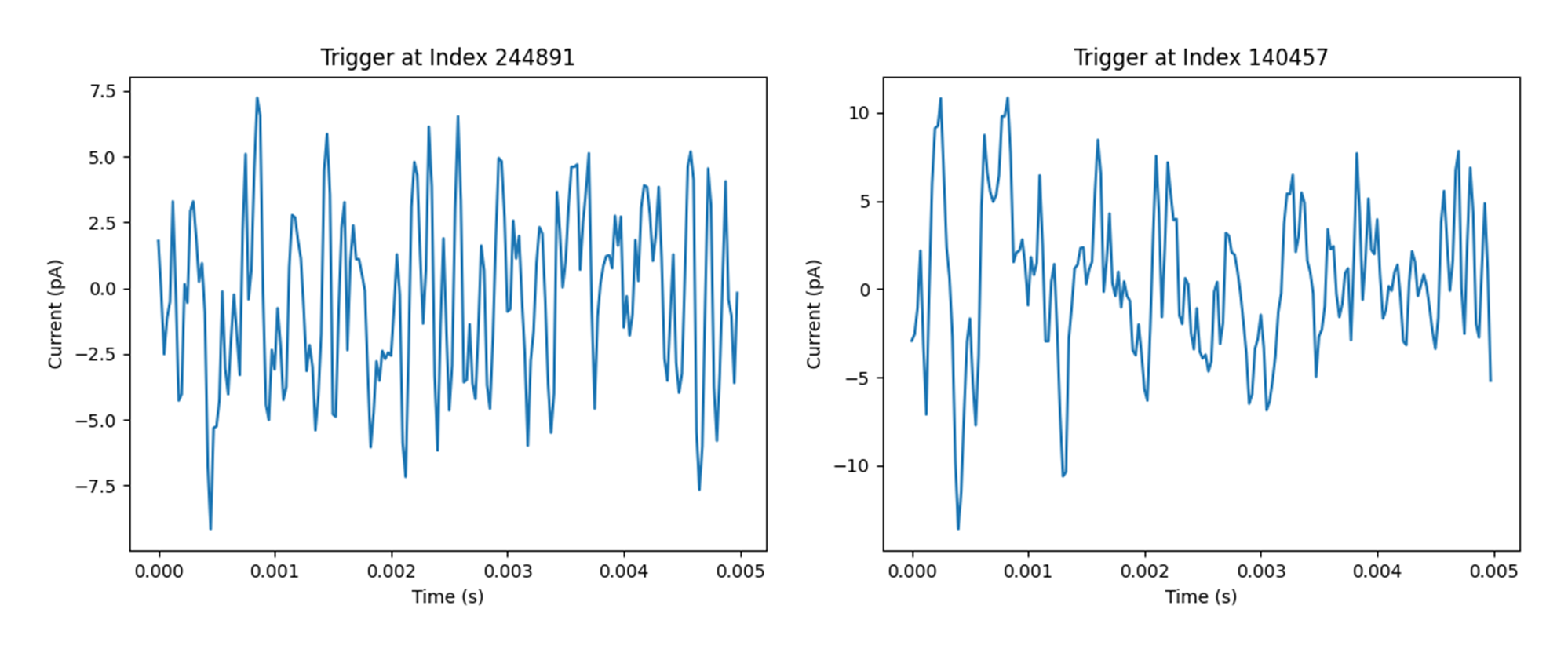}\caption{\label{fig:Noise-masks-collected}Examples of noise masks collected
from the recordings of cell A140724\#065 (\emph{on the left}) and
A160127\#015 (\emph{on the right}).}
\end{figure}

The synthetic data consisted in 12,700,600 action potentials samples of 160 points (simulating 4 ms at 40 kHz of sampling), 6,675,300 of which emulated action potentials from serotonergic cells and 6,025,300 from non-serotonergic action potentials. From the original training data recordings we extracted 600 noise masks (see e. g. Fig. \ref{fig:Noise-masks-collected}) from a selection of which were randomly applied to the biological action potentials thus obtaining the synthetic data (see e. g. Fig. \ref{fig:SyntheticData}). The purpose of generating the synthetic data, besides plain data augmentation for higher accuracy, is also to provide an estimate of the overfitting of the biological model based on the noise signature of the data. 

The generation of the synthetic data was done according the following
procedure. Each original training data sample is smoothed through
averaging, i.e. the values of the smoothed sample $\left\{ y'_{m}\right\} $
with $m\in\left\{ 1,...,160\right\} $, are given as the averages
of the values of the original sample $\left\{ y_{m}\right\} $ by
\begin{equation}
y'_{m}=\frac{\left(y_{m-1}+y{}_{m}+y{}_{m+1}\right)}{3}.
\end{equation}
The reason for this 3-point averaging preprocessing of the signal is due to the need to combine two requirements: the need to smooth the original signal from the specific noise of the recording, and the need to maintain the structure of the signal. The rapid depolarization of the cell is such that the most relevant data of the action potential recording are often formed in a few tenths of a millisecond, i.e., most useful informations are supposedly condensed in about a dozen of recording points. This means that considering $n$-point averaging with $n>3$ could undermine the fundamental information inside the signal, while $n=2$ might not be sufficient to remove the background noise.
After the averaging process, the values of smoothed action potential $\left\{ y'_{m}\right\} $ are added to the values of randomly chosen noise mask $\left\{ n_{m}^{(k)}\right\} $
where $k\in\left\{ 1,...,600\right\} $ is randomly chosen. The final
synthetic sample is thus obtained as the sample $\left\{ y^{(k)}{}_{m}\right\} $
with
\begin{equation}
y^{(k)}{}_{m}=y'_{m}+\alpha \cdot n_{m}^{(k)},
\end{equation}
where $\alpha\in[ 0.2,0.4]$ is a randomly generated “dumping coefficient”  experimentally found around $0.3$ to modulate the noise. The choice of this coefficient requires some clarification. Indeed, the coefficient dumps the noise intensity to synthesize more physiologically plausible spike waveforms. First, the background noise was not completely removed when averaging action potentials, just smoothed with a 3-point average. Thus directly adding the full noise mask would excessively boost the background noise compared to the original recording. Moreover, the original noise does not influence all points of the signal equally, but is more pronounced in slower changing current regions. Applying the raw noise mask tends to produce unrealistic action potential shapes, e.g. double bottoms. The dumping coefficient between $0.2$ and $0.4$ was deemed a suitable range by visual inspections by an expert author with over 30 years of experience on serotonergic action potential recordings. 

\begin{center}
\begin{figure}
\begin{centering}
\includegraphics[scale=0.70]{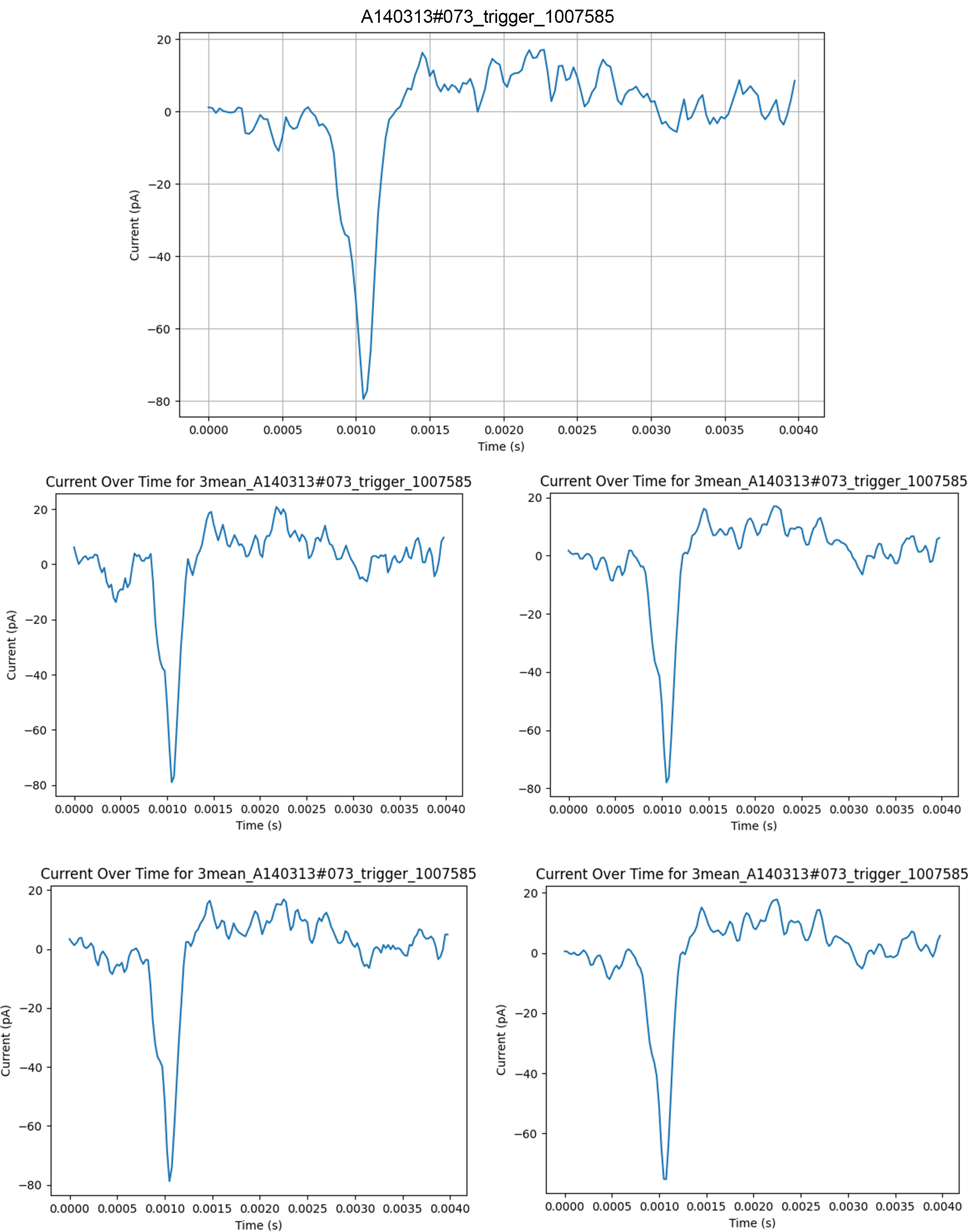}
\par\end{centering}
\caption{\label{fig:SyntheticData}Example of 4 synthetic action potentials generated
by the event triggered at 1007585, i.e. 25.189 sec, of the serotonergic
cell A140313\#073. Top trace: the original recording of the event.
The panels report four action potential obtained by processing the original trace
with different noise masks (see methods).}
\end{figure}
\par\end{center}

\subsection{Model Description}

In accordance with the origin of our dataset, we developed two distinct models, namely the ``biological'' model (trained only over original
data) and the ``synthetic'' model (trained only on synthetic data).
The biological model underwent training, validation, and testing using
the original training data, which comprises 27,108 action potential samples after the balancing of the classes.
Conversely, the synthetic model was trained, validated, and tested
utilizing synthetic data, encompassing 12,700,600 action potential samples.  

Fig. 1 summarizes the various steps used to implement the model from
recorded signals. The architecture of the models is a sequence of
layers commonly used in deep learning, specifically in the context
of convolutional neural networks (CNNs) for image or signal processing.
We implemented the architecture using the Keras libraries in TensorFlow
2. The model of the neural network consists of a normalization layer
for stabilizing the learning process and reducing training time; two
repetitions of a 2D convolutional layer with 32 filters and a max
pooling layer with a pool size of (2x1); a flatten layer to connect
to a dropout layer and dense layers with 2 output units used for binary
classification. Activation functions of the convolutional layers are
the ReLU, while for the dense layer we used the classic sigmoid (see
Table 1 for a summary of the model). 
\begin{table} 
\centering{}%
\begin{tabular}{ccc}
\emph{Layer (type)} & \emph{Output Shape} & \emph{Param \#}\tabularnewline
\hline 
Layer Normalization & (None, 160, 2, 1) & 320\tabularnewline
Conv2D & (None, 141, 2, 32) & 672\tabularnewline
MaxPooling2D & (None, 70, 2, 32) & 0\tabularnewline
Conv2D & (None, 51, 2, 64) & 41024\tabularnewline
MaxPooling2D & (None, 25, 1, 64) & 0\tabularnewline
Flatten & (None, 1600) & 0\tabularnewline
Dropout & (None, 1600) & 0\tabularnewline
Dense & (None, 2) & 3202\tabularnewline
\hline 
\emph{Total Params} & 45218 & \tabularnewline
\end{tabular}\caption{Summary of the CNN architectural model with kernel 20 used for the
neural network. Other models follow the same architectural structure
and change only for the dimension of the kernel.}
\end{table}
For training we chose the "binary crossentropy"
loss function, which is standard for binary classification problems,
while the optimizer was "Adam" (Adaptive Moment Estimation) as
these are common choices. A special treatment was devoted to the kernel
of the 2D convolutional layers. Indeed, since the kernel of these
layers express the ability of the convolutional process in enlarging
a specific portion of the pattern, we explored a range of possible
kernels between 1 to 31.  All models were trained on 25 epochs with a batch size of 64 and their test accuracy ranged from 88.3\% (model with kernel 1) to 98.4\%
(model with kernel 23) with a test loss of 0.2641 and 0.05524.
To enhance the robustness of the model, instead of selecting a single
kernel and using one model for inference, we selected all models with
kernels ranging between 20 and 30 and took the consensus between the
models. This technique ensures more stability in the overall architecture
and is often considered best practice.
Since this article presents a method rather than offering a specific optimized deep learning model, we did not systematically search for a specific architecture other than the one above which is a standard. However, we explored a few different architectures with varying numbers of layers and neurons per layer.  Nevertheless, the improvement in accuracy was not enough to justify adopting a more complicated architecture. At this stage, to our understanding, acquiring more data represents the most relevant advancement for achieving a better model. Nevertheless, since this article is just a proof of concept, we leave open the possibility of future research into the most suitable architecture for this problem.

Finally, it is worth noticing that while the training of the biological
model did not require  any specific adjustment, the synthetic model,
involving $>$ 12M action potential samples required a continuous learning implementation,
where the model was trained over 200 training sessions of 63,450 synthetic
action potential samples.

\subsection{Assessment of Accuracy and Sensitivity}

For the assessment of the models we used the following metrics: \emph{Accuracy},
\emph{Sensitivity at Specificity 0.5}, \emph{Area Under the Curve (AUC)}, \emph{F1-Score} and the \emph{Confusion Matrix}. 
\begin{itemize}
\item Accuracy measures the proportion of total predictions (both serotonergic
and non-serotonergic cells) that the model correctly identifies, i.e.
\begin{equation}
Accuracy=\left(\text{True Positives}+\text{True Negatives}\right)/\text{Total Samples}.
\end{equation}
This metric was chosen for identifying if the models are generally
effective in classifying both serotonergic and non-serotonergic cells. 
\item Sensitivity at Specificity measures the sensitivity of the model,
i.e.
\begin{equation}
Sensitivity=\text{\text{True Positives}}/\text{\ensuremath{\left(\text{True Positives}+\text{False Negatives}\right)}},
\end{equation}
at a fixed specificity, i.e. $\text{\text{True Negatives}}/\text{\ensuremath{\left(\text{True Negatives}+\text{False Positives}\right)}}$,
which we set at 0.5 . The choice of this metric with this setting
ensures that the models are not overly biased towards identifying
serotonergic cells at the expense of misclassifying non-serotonergic
ones. 
\item  The \emph{Area Under the Curve} (AUC) of the Receiver Operating Characteristic (ROC) provides a measure of the model's ability across classification thresholds, i.e.,
\begin{equation}
\text{AUC} = \int_{0}^{1} \text{TPR}(t) , dt,
\end{equation}
where $t$ is the threshold, and $TPR(t)$ is the true positive rate at threshold 
$t$. This metric is particularly useful because it is independent of the classification threshold and provides a single measure of performance across all possible levels of sensitivity and specificity. 

\item  The \emph{F1-Score} is the harmonic mean of precision and sensitivity (recall).
The F1-Score takes both false positives and false negatives into account, providing a  balanced view of the model's performance. We considered useful this metric for the measuring the robustness of the model, balancing the trade-off between precision and recall. 
\item Finally, the Confusion Matrix shows the percentages of True Positives,
False Positives, True Negatives, and False Negatives giving a complete
feedback of the models. This is a a detailed view that we considered
essential for understanding the specific areas where the models need
improvements.
\end{itemize}
All these metrics were used for all the data, i.e. Original Training
Data, Non-homogeneous Data and Synthetic Data. In the specific case
of the Original Training Data all the metrics were used in the three
phases of Training, Validation and Testing. The training phase was
developed on 30,328 action potentials selected uniquely for training. The Validation
phase, which is used to tune hyperparameters, was on 6,500 action potentials
which the model has not seen during training. Finally, the last 6499
action potentials were used for the Testing of the models, and are those on which
the true performance of the models is assessed.

\subsection{Repository of the Model and Data }

We made available in the GitHub respository at \texttt{GitHub.com/neuraldl/ DLAtypicalSerotoninergicCells.git} 
the following: 
\begin{enumerate}
\item the .abf recordings of original training data and the non-homogeneous
data,
\item the 43,327 single action potentials samples of the original training data stored in .csv files of 160 points,
\item the 24,616 single action potentials samples of the non-homogeneous data stored
in .csv files of 160 points,
\item the 12,700,600 million single action potentials samples of the synthetic data
stored as numpy vector,
\item the trained models with different kernels,
\item the results of the models,
\item the Python notebooks for training of the models and for inference.
\end{enumerate}

\section{Results}

In this study we compared the spiking activity of 300 neurons recorded
in DRN slices obtained from transgenic mouse lines with serotonergic
system-specific fluorescent protein expression. 

\subsection{Visual discrimination of action potentials}

\begin{figure}
\includegraphics[scale=0.8]{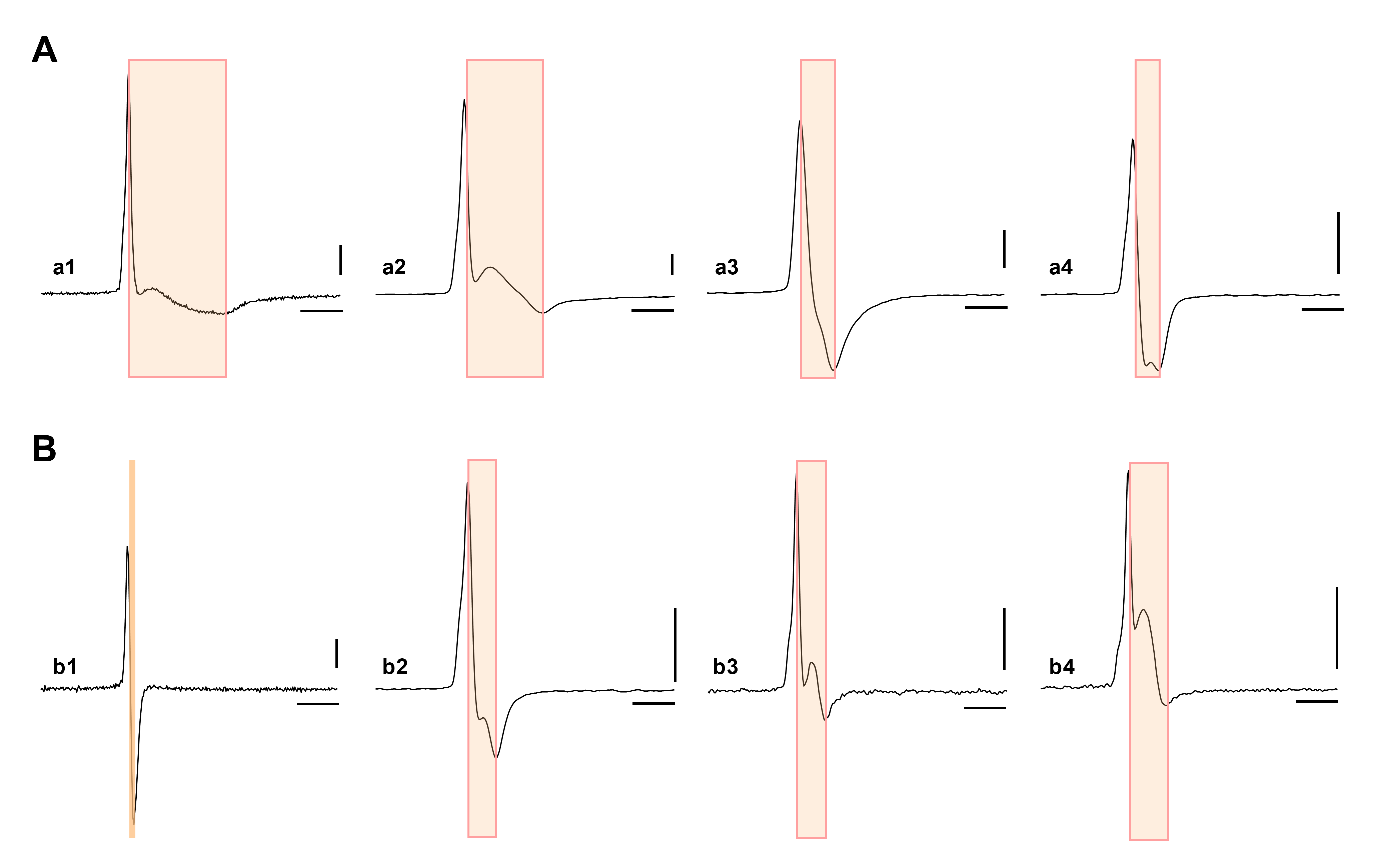}\caption{\label{fig:Examples-of-spikes}Examples of action potentials recorded from serotonergic
and non-serotonergic neurons in slices of dorsal raphe nucleus. A.
Fluorescent protein-labelled (serotonergic) neurons: a1,a2: typical
action potentials of serotonergic neurons; note the long interval between spike
upstroke and downstroke (UDI) highlighted by the shaded area in all
traces. a3-a4: recordings from serotonergic neurons displaying spikes
of shorter duration. B. Fluorescent protein-unlabelled (non-serotonergic)
neurons: b1: typical biphasic spike of short duration from
a non-serotonergic neuron; b2-b4: spikes of variable shape
recorded from non-serotonergic neurons. Shaded areas indicate the
width of the spike measured by UDI (see methods). Note the overlap
in action potential width of some serotonergic and non-serotonergic neurons.
Traces are averages of 15-50 sweeps. Calibrations 25 pA (polarity
inverted); 1 ms.}
\end{figure}

As illustrated in Fig. \ref{fig:Examples-of-spikes}, serotonergic
neurons displayed action potentials of different shape and duration that were
often difficult to be discriminated from those observed in non-serotonergic
neurons. Thus, with the exception of action potentials showing the typical shape
and duration of serotonergic neurons (e.g. Fig. \ref{fig:Examples-of-spikes}:
traces a1, a2) or of non-serotonergic cells (e.g. Fig. 6: trace b1)
both types of neurons may display action potentials similar in width and/or shape.
Therefore, the sole duration of the spike, which could be determined
online by measuring the upstroke/downstroke interval (UDI) may result
not conclusive for immediate serotonergic neuron identification.

\begin{figure}
\includegraphics[scale=0.7]{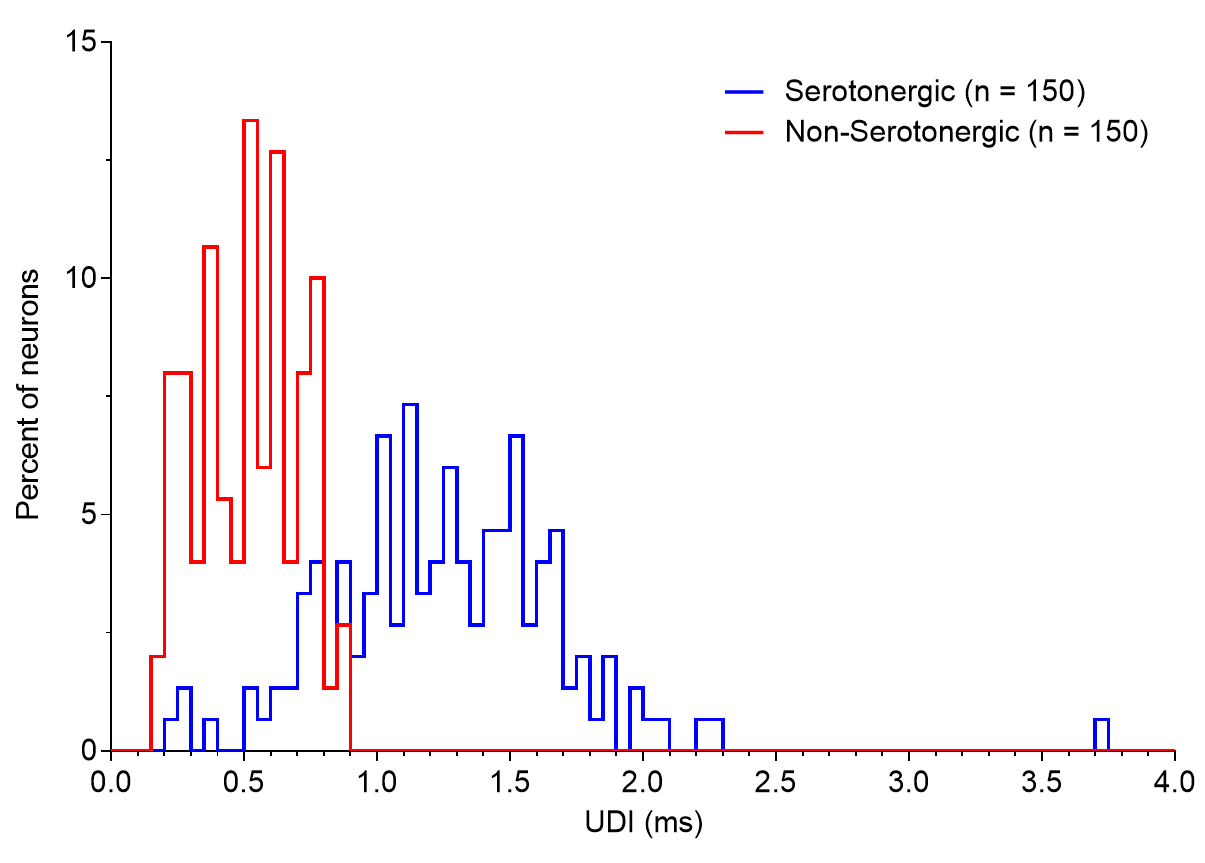}\caption{\label{fig:Distribution-of-spike}Distribution of spike duration of
serotonergic and non-serotonergic neurons. Histograms report the distribution
of spike width measured by the interval between spike upstroke and
downstroke (UDI) in serotonergic (blue) and non-serotonergic neurons
(red) recorded in slices of dorsal raphe nucleus. Note the overlap
in spike duration between serotonergic and non-serotonergic neurons. Firing rate (mean ± s.e.m) of serotonergic and non-serotonergic neurons was 2.17 ± 0.08 Hz (range 0.51-5.80 Hz; n=150) and 3.6 ± 0.32 Hz (range 0.07-16.60 Hz; n=150), respectively.}

\end{figure}

From our database of recordings we have selected 150 serotonergic
neurons labelled by fuorescent proteins and 150 non labelled cells,
deemed to be non-serotonergic cells. The distribution of spike
width of these two populations is shown in Fig. \ref{fig:Distribution-of-spike}.
 These neurons were chosen on the sole technical characteristic of
not showing detectable artefactual transients that could be mistaken
by the deep-learning routine as action potentials. From these two
populations of neurons we extracted 108 serotonergic and 45 non-serotonergic
neurons to implement the training of the Biological Model. In addition,
12 serotonergic neurons from three different experimental days and
10 non-serotonergic neurons from four different experimental days
were used for testing the model with data non homogeneous to the training
(see methods). 

\begin{figure}
\includegraphics[scale=0.6]{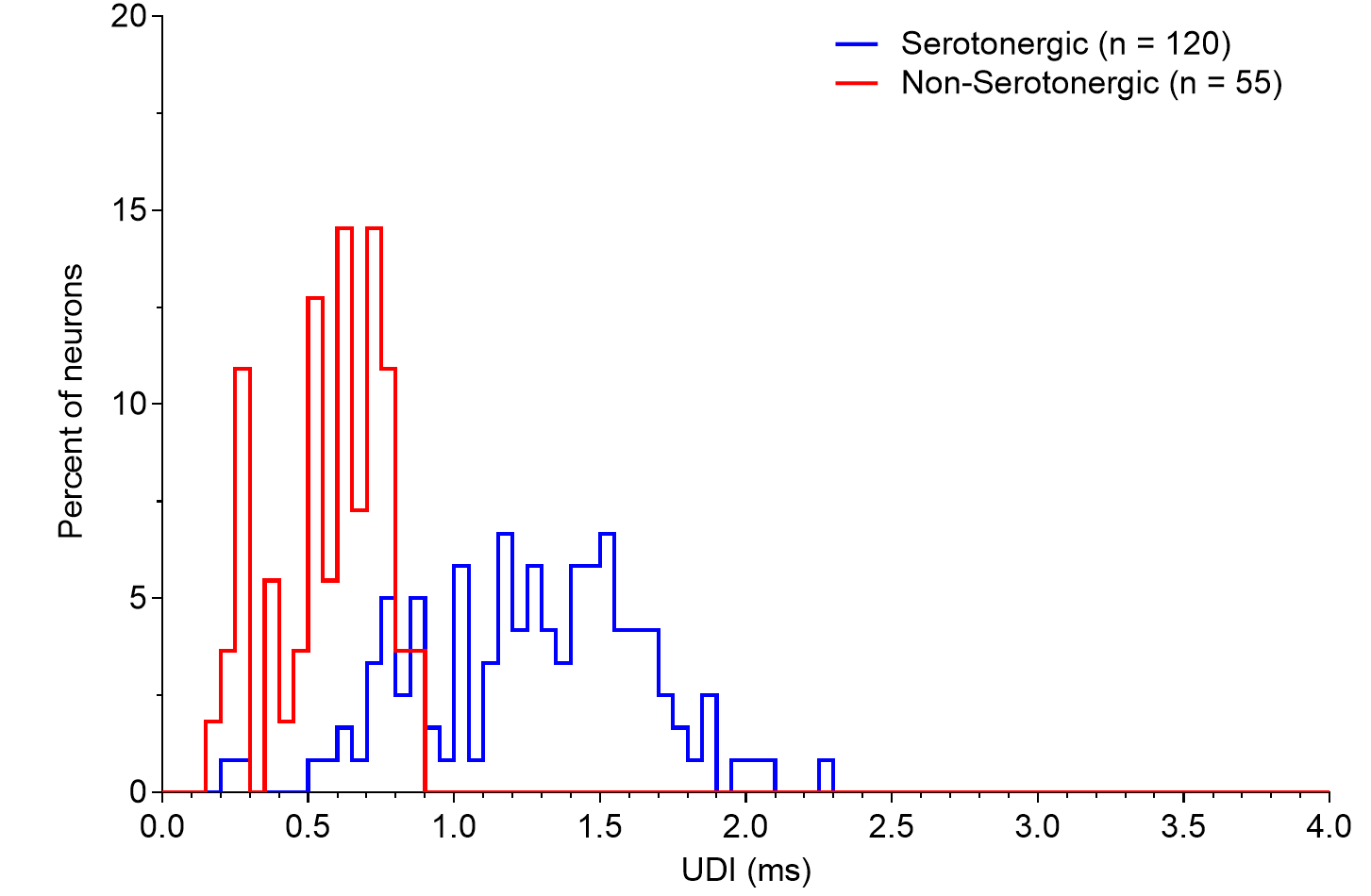}

\caption{\label{fig:Distribution-of-spike-1}Distribution of spike duration
recorded from the neurons utilized to develop the model 1. Histograms
report the distribution of spike width measured by the interval between
spike upstroke and downstroke (UDI) in serotonergic (blue) and non-serotonergic
neurons (red) recorded in slices of dorsal raphe nucleus. Firing rate (mean ± s.e.m) of serotonergic and non-serotonergic neurons was 2.29 ± 0.09 Hz (range 0.38-5.08 Hz; n=120) and 2.43 ± 0.33 Hz (range 0.07-9.32 Hz; n=55), respectively.}

\end{figure}

As shown in Fig. \ref{fig:Distribution-of-spike-1}, the neurons used
from training and testing the model 1 are representative of the two
(serotonergic and non-serotonergic) populations of neurons.

An additional group of recordings ($n=30$) from fluorescence identified
serotonergic and non-serotonergic neurons, not previously used for
the model implementation, were processed by the model 1 to test its
ability to recognize cell type from the spike characteristics distilled
by the model itself.

\subsection{Discrimination with Deep Learning Models}

The metrics of both the biological model and the synthetic model were
collected over the testing data (original and synthetic) during their
training phase, as standard practice in deep learning. Over this data
both the biological model and the synthetic model scored $>$ 98\%
accuracy. In addition to the standard practice we evaluated the models
over non-homogeneous data in order to evaluate possible sources of
overfitting arising from noise signatures in the recordings.  On this
dataset the biological model scored $>$ 91.2\% accuracy showing
the existence of some light source of overfitting. As expected the synthetic model showed better results with $>$ 93.7\% accuracy. Overall, we consider the metrics evaluated on non-homogeneous data more indicative and more reliable than those arising from the training data. Indeed, non-homogeneous data not only were unknown to the model, but were also collected on different days than those of the data used for the training. 

\subsubsection{Results on the Training Data }

\paragraph{Biological Model}

The biological models, when tested on the original training dataset, showed varying performance metrics. For kernels ranging from 1 to
31, the test loss was observed between 0.26417 (kernel 1) and 0.05006
(kernel 27). Accuracy measurements ranged from 0.88296 (kernel 1) to 0.98401 (kernel 29), as detailed in Fig.
\ref{fig:Test-loss,-accuracy,}. The consensus biological model, obtained from models with kernels 20 to 30, tested on the original data recorded a test loss of 0.05457, an accuracy of 0.98401, and a sensitivity at specificity 0.5 of 0.99852, an AUC of 0.99747 and an F1-Score of 0.98340 as shown in the last row of Table \ref{tab:TabBiological}). 
\begin{table}
\begin{centering}
\begin{tabular}{c|ccccc}
\emph{Kernel} & \emph{Test Loss} & \emph{Accuracy} & \emph{Sens. at Spec. 0.5 }& \emph{AUC }& \emph{F1-Score }\tabularnewline
\hline 
1 & 0.26417 &	0.88296 &	0.99409 &	0.95765 & 0.88333\tabularnewline
5 & 0.10187 &	0.95893 &	0.99901 &	0.99314 & 0.95941 \tabularnewline
10 & 0.06001 &	0.97934 &	0.99926 &	0.99729 & 0.97910 \tabularnewline
15 & 0.06673 &	0.98155 &	0.99827 &	0.99663 & 0.98143 \tabularnewline
20 & 0.05831 &	0.98303 &	0.99852 &	0.99706 & 0.98278 \tabularnewline
25 & 0.05343 &	0.98131 &	0.99901 &	0.99789 & 0.98119 \tabularnewline
30 & 0.07726 & 0.97713 &	0.99704&	0.99544 & 0.97726 \tabularnewline
\hline 
\emph{Biological model} & 0.05457 &	0.98401 &	0.99852 &	0.99747 & 0.98340 \tabularnewline
\end{tabular}
\par\end{centering}
\begin{centering}
\caption{\label{tab:TabBiological}A selection of the metrics on the test data of the models trained with Original Training Data. Beside \emph{Accuracy}, \emph{Sensitivity At Specificity 0.5}, \emph{AUC} and \emph{F1-Score} we reported also \emph{Test
loss}, which represent the error between the predicted values and the actual values and is a standard metric in evaluating DL models. Values reported in last row ``\emph{biological model}'' refer to the metrics of the full biological model given as the consensus of
the single models with 20 $\protect\leq$ kernel $\protect\leq$ 30.}
\par\end{centering}
\end{table}
\begin{figure}
\centering{}\includegraphics[scale=0.8]{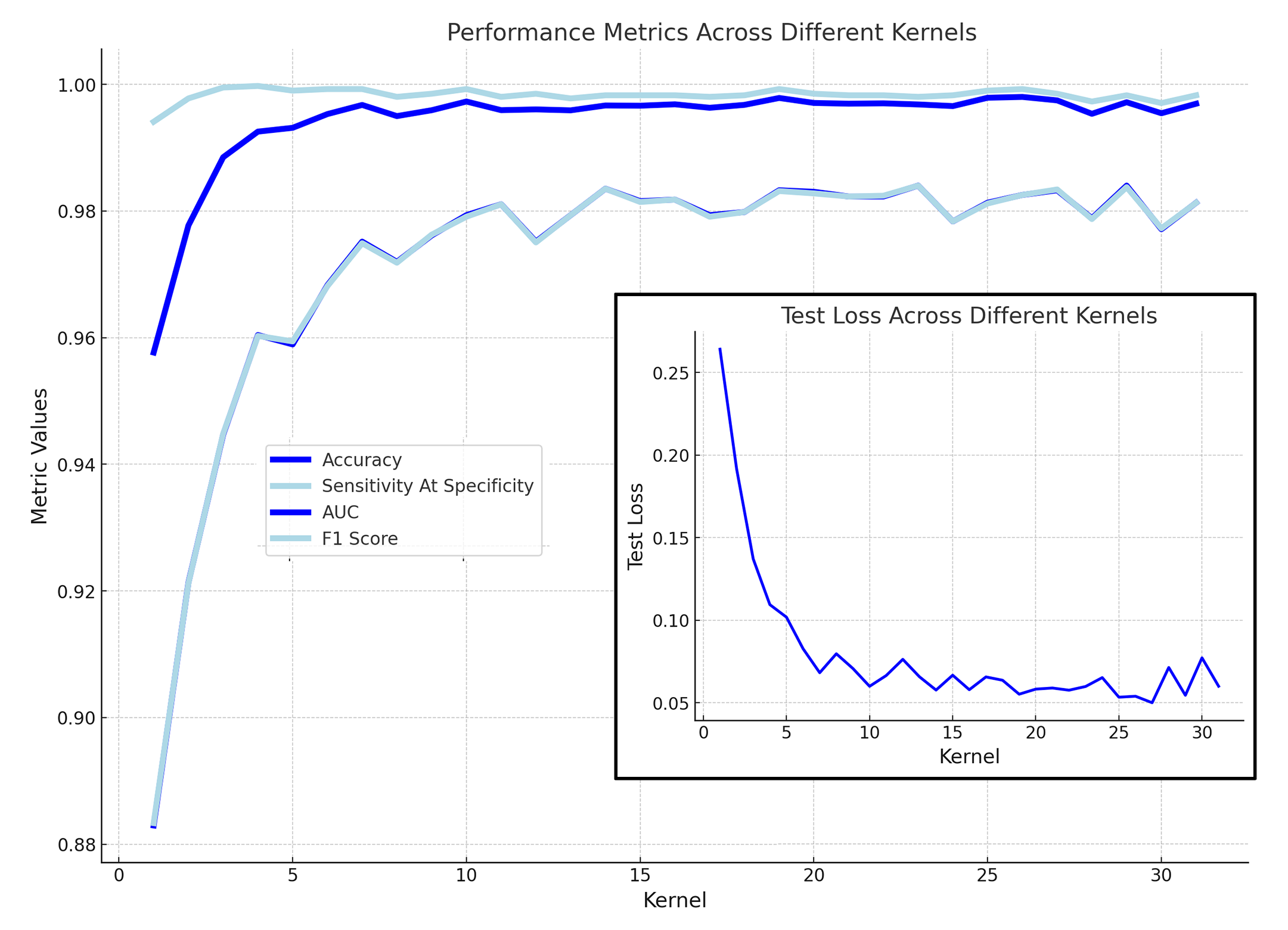}\caption{\label{fig:Test-loss,-accuracy,}Each of the 32 models, with kernel
sizes varying from 1 to 31, was evaluated for test loss, accuracy, sensitivity at a specificity of 0.5, AUC and F1-Score. The resulting graphs depict a monotonic trend correlating with the increasing kernel sizes, which
eventually stabilizes in the range from kernel size 20 to 31.}
\end{figure}

\paragraph{Synthetic Model}

\begin{figure}
\centering{}
\includegraphics[scale=0.60]{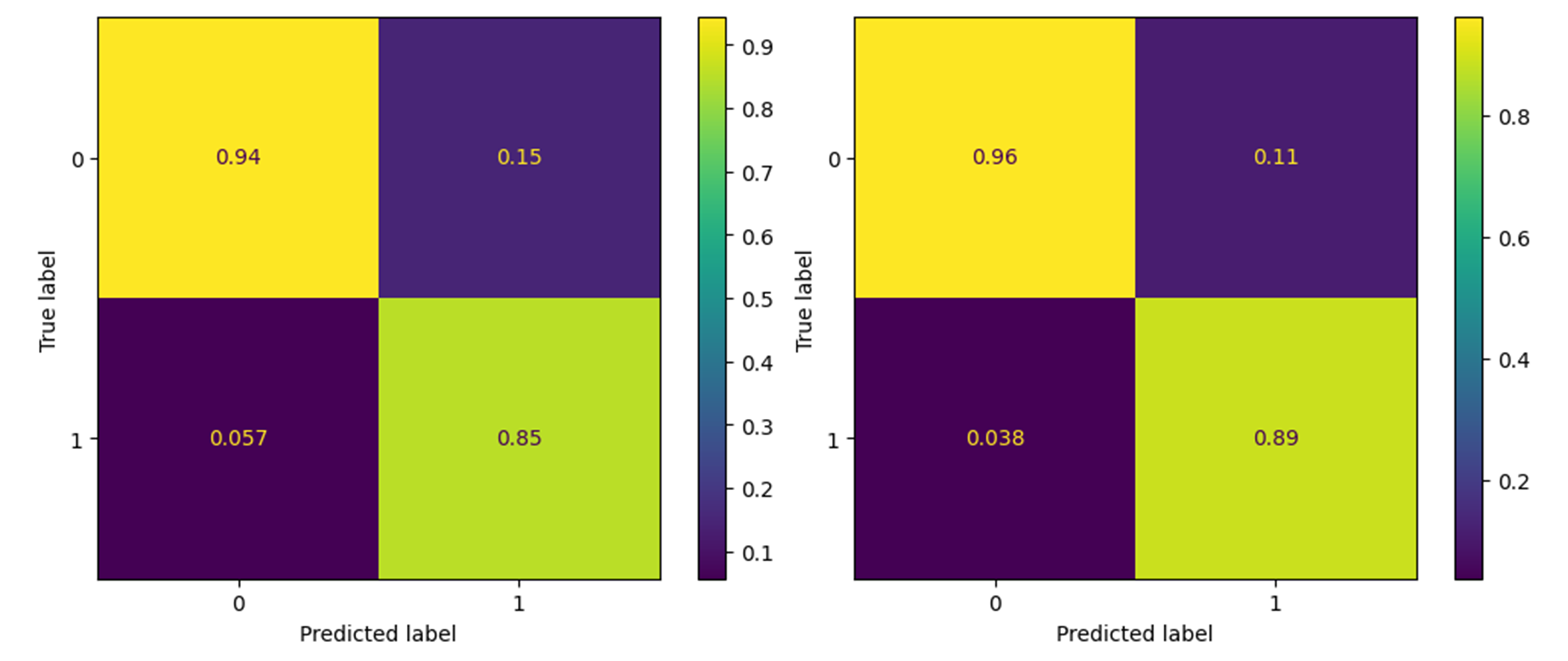}

\caption{\label{fig:Confusion-matrix-of} Confusion matrix for the biological
and synthetic model over the non-homogeneous data labels serotonergic cells as 0
and non-serotonergic cells as 1. The matrix of the biological model (lest panel) shows the True Positive Rate 94.4\% at the Top-Left; the False Negative Rate 5.6\% at the
Bottom-Left; the False Positive Rate 14.8\% at the Top-Right; and the
True Negative Rate 85.1\% at the Bottom-Right. On the other hand,  the matrix of the synthetic model (right panel) shows the True Positive Rate 96.2\% at the Top-Left; the False Negative Rate 3.7\% at the
Bottom-Left; the False Positive Rate 11.1\% at the Top-Right; and the
True Negative Rate 88.8\% at the Bottom-Right. In the table below all accuracy types of the two models over non-homogeneous data} 
\bigskip
\begin{tabular}{c|cccc}
\emph{Model}  & \emph{Accuracy} & \emph{Sens. at Spec. 0.5 }& \emph{AUC }& \emph{F1-Score }\tabularnewline
\hline 
\emph{Biological model} & 0.9125 & 0.8518 & 0.8976 & 0.8679\tabularnewline
\hline 
\emph{Synthetic model}  & 0.9375 & 0.8888 & 0.9255 & 0.9056\tabularnewline
\end{tabular}
\end{figure}

The evaluation of the 32 synthetic models on the synthetic dataset yielded superior metrics compared to the biological models on the training dataset. These results on the training dataset are not deemed highly significant, as overfitting not related to recording noise tends to be amplified in the augmented dataset. However, we considered significant the results of the synthetic model on non-homogeneous data. Indeed, as pointed out in Fig. \ref{fig:Confusion-matrix-of} the synthetic model outperformed the biological model on non-homogeneous data. 

\subsubsection{Results on Non-homogeneous Data }

The most significant outcomes were derived from non-homogeneous data,
i.e., cells that were not utilized in training and that were collected on different days other than those used for the training data. Using this dataset, the
biological model achieved an accuracy of 0.9125, a sensitivity at specificity of 0.5 of 0.8518, an AUC of 0.8976 and an F1-Score of 0.8679. An even better result was given by the synthetic model which achieved an accuracy of 0.9375, a sensitivity at specificity of 0.8888, an AUC of  0.9255 and an F1-Score of 0.9056.
A crucial indicator of performance
is the  confusion matrix (refer to Fig. \ref{fig:Confusion-matrix-of}). The best results were obtained by the synthetic model. Indeed, out of 55 serotonergic cells, 53 (96.2\%) were accurately identified as serotonergic (True Positive), while 2 (3.8\%) were incorrectly classified as non-serotonergic (False Negative). Conversely, of the 27 non-serotonergic cells, 24 (88.8\%) were correctly recognized (True
Negative), and 3 (11.1\%) were erroneously labeled as serotonergic (False Positive). The biological model had similar results but misclassified 3 (5.5\%) serotonergic cells as non-serotonergic an 4 (14.8\%) non-serotonergic cell as serotonergic.  In the non-homogeneous data the False Positive Rate is  higher than the False Negative Rate. We do not have explanation for this phenomenon other than randomness. Indeed, in the original data, the False Positive and False Negative Rates are similar, and this phenomenon is not present when testing either the biological or synthetic models.
\begin{figure}
\centering{}\includegraphics[scale=0.62]{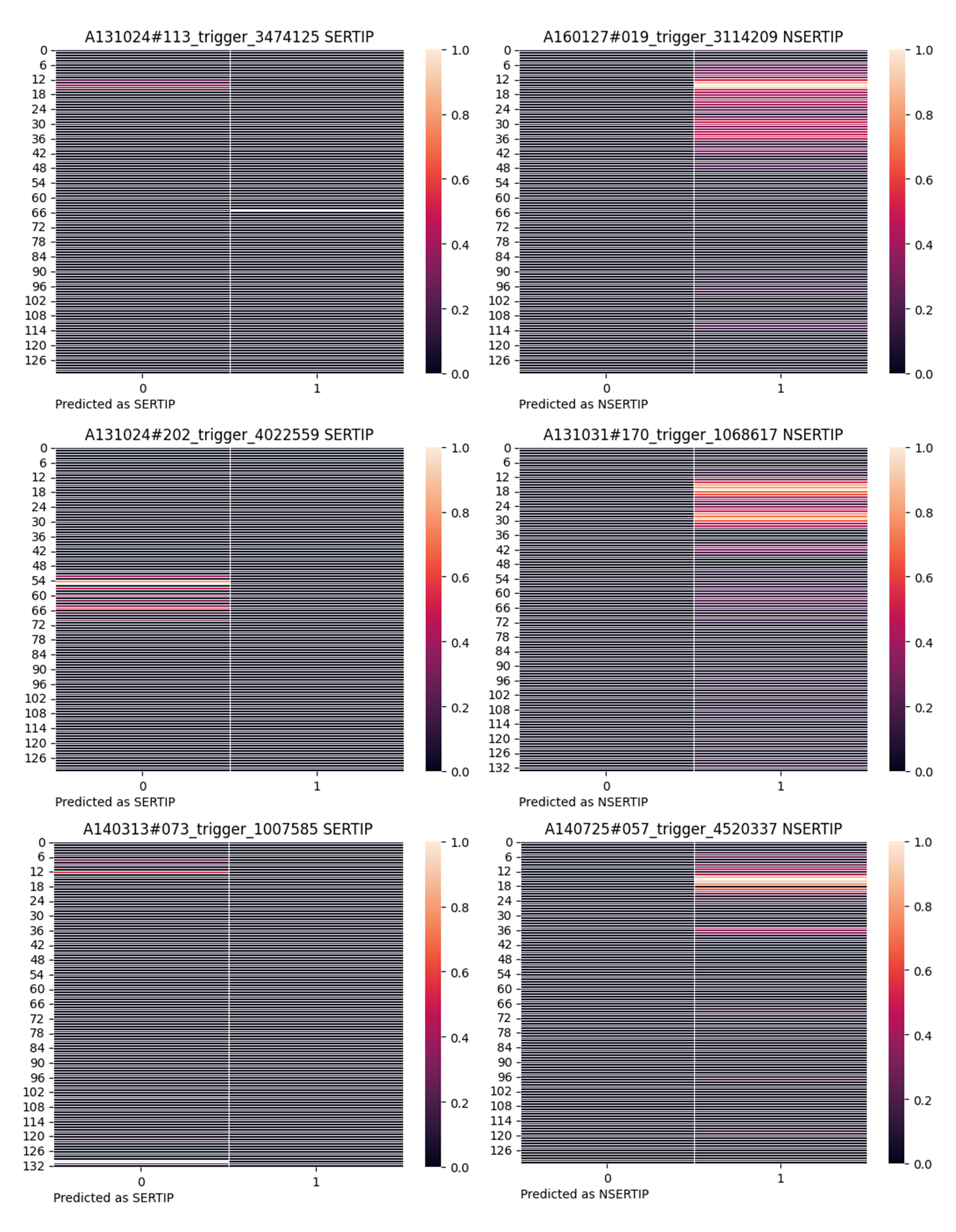}\caption{\label{fig:CamGrad} The above images are Gradient-weighted Class Activation Mapping (Grad-CAM) visualisations that highlight the regions of the input signal deemed most significant by the first convolutional layer (Conv2D) of the biological model for the classification of three serotonergic and three non-serotonergic cells. Although not a definitive pattern, the model tends to focus more on the initial portion of the signal, particularly the spike. In each panel, the left scale indicates the time step of the input signal, while the right colour scale shows the activation intensity within the Conv2D layer's neurons.} 
\end{figure}

\section{Discussion}

Deep-learning based models have gained increasing importance in biomedicine for their high performance in image processing and morphological recognition
of cells that can be applied both in clinical diagnostics (Johansen
et al., 2016; Litjens et al., 2017; Rácz et al., 2020) and in preclinical
research when complex patterns of data need to be measured, classified
and interpreted (De Luca et al., 2023). More specifically, convolutional
neural networks (CNN) effectively address complex pattern recognition
especially when patterns are hidden across varying scales and orders
of magnitude. This is highly relevant in neuronal action potentials, where the
peak impulse and the rise of the spike may occur in a fraction of
millisecond, whereas the interval between spikes can be vastly longer.
The here proposed model provides an important proof of concept for
usefulness of CNN for identification of neuron types in the central
nervous system on the basis of their spiking activity. To the best
of our knowledge, this is the first time that this type of architecture
is applied to recognition of neuronal action potentials by their recorded traces.
Moreover, the recognition of serotonergic neurons has been validated
by an independent identification of the recorded neuron by its serotonin
neuron specific expression of a fluorescent protein.

\subsection{Comparison with existing procedures for serotonergic neuron identification
from their physiological activity}

Identification of different neurons active in a restricted brain area on the basis of their spike shapes may be a valid and sufficient criterion when the characteristics of spikes can reliably be separated in different classes. For instance, Tseng and Han (2021) recorded \emph{in vivo} the activity of behavioural-task responsive neurons of prefrontal cortex in mice and discriminated excitatory and inhibitory neurons taking advantage of the known, clearcut, difference in the duration of spikes in the two classes of neurons. In contrast, our DL based model finds its application when the characteristics of spikes from different neurons overlap as for serotonergic and non-serotonergic neurons of the dorsal raphe nucleus. Indeed, automatic routines for online measurements of action potentials can be designed, however until now no valid criteria for discriminating between spikes generated by serotonergic and non-serotonergic neurons have been established. Recognition of serotonergic neurons during
extracellular recordings relies mostly on visual evaluation of the
shape of the spike, that is often polyphasic, combined with the regular
firing activity at relatively low frequency (up to 3-4 Hz). Thus,
the mean criterion is mainly based on the asymmetric proportion between
the upstroke and the downstroke of the spike (with a ratio usually
$>$2.5) and duration of the spike itself after the main upstroke (usually
accepted in the range $>$ 1.2 ms). Coexistence of these characteristics
is sufficient to enable an experienced Researcher to identify typical
serotonergic neurons with a high degree of confidence. Nevertheless,
in our recordings from genetically identified serotonergic neurons
we noticed relatively frequent deviations from these criteria. Indeed,
the spike duration of several neurons was less than 0.9 ms, down to
0.4-0.5 ms (see Fig. 6 in Mlinar et al., 2016). Similarly, a not negligible
percentage of non-serotonergic neurons displayed spike shape and firing
characteristics different from the expected biphasic, symmetric spikes
of brief duration ($<$0.5-0.6 ms) and high frequency, often irregular,
firing. Thus, some non-serotonergic neurons show long and asymmetric
action potentials and sometimes have a regular, low frequency activity
($<$4 Hz) which makes their recognition difficult. Therefore, while
\textquotedblleft typical\textquotedblright{} serotonergic and non-serotonergic
neurons are relatively easy to be discriminated with the currently
accepted criteria a number of serotonergic neurons that do not comply
with the classically established recognition criteria are discarded
and not studied for their pharmacological and physiological characteristics. 

Given these limits of the online visual recognition of serotonergic
neurons, our model provides a valid tool for the intra-experiment
identification of neurons recorded in the dorsal raphe nucleus, as
the model can be implemented in the initial routine of \emph{in vitro}
recordings. Notably, our DL biological model relies only on spike
shape for recognition of serotonin neurons and therefore it enables
the identification and investigation of subpopulations of serotonergic
neurons displaying irregular firing or low frequency oscillatory patterns
of firing (see e.g. Mlinar et al., 2016). 

\subsection{Characteristics and limits of the model for its application }
For the identification of serotonergic neurons we relied on transgenic mice lines that express fluorescent marker proteins under the control of serotonergic system-specific Tph2 and Pet-1 promoters. While the Tph2 promoter-driven expression of fluorescent reporter genes is expected to unmistakably label serotonergic neurons, there is a possibility that the Pet-1 Cre-based method does not label few serotonergic neurons in the DRN (see in Mlinar et al., 2016 for more detailed discussion and references). In the present context we used both promoters and the probability that these rare neurons participated in the training of the model is very low. In this unlikely case, as these non fluorescent cells were not categorized as serotonergic neurons during the model training, this specific serotonergic neuron subtype would probably be misclassified by the model.

It should be mentioned that the present model applies to the specific
recording method used in collecting our database of action potentials. Thus,
for immediate application of the model the sampling frequency should
be set at 40 kHz. Our data were acquired using the Clampex program
in loose-seal cell attached patch clamp mode, but since the routine
transforms the signals in {*}.csv files any acquisition program that
produces files in a format that can be transformed in {*}.csv format
would provide adequate input for the model. The amplitude of the recorded
current should be greater than the detection threshold that we have
imposed in the model to minimize acquisition of small transients ($>$50
pA). Finally, our recordings were performed at the temperature of
\textasciitilde 37 °C. Although small deviations from this temperature
could be tolerated, it should be considered that the width of the
action potential, may be influenced by temperature. Notwithstanding these limitations,
if the sampling rate is adequate and the signal reaches the detection
amplitude the model provides an answer on neuron type with an accuracy
of $>$91.2\% within an inference time of a few milliseconds after the
submission of the recorded traces (the inference time is the raw time
taken by the model in classifying the signal without considering the
latent time of converting and transmitting the signal to the model
which can vary depending on the user interface chosen in the deployment
of the model).

It is noteworthy that in several experiments ( $\sim30\%$ of those
used here) used for training the model we applied a gentle suction
in the patch pipette during the recording to improve the signal to
noise ratio. We have previously shown (Mlinar et al., 2016) that this
procedure does not alter the shape and duration of the recorded signals.
In our context, this intra-experiment change in the amplitude of events
recorded from the same neuron increases the robustness of the training
data because implemented the model with events of constant shape but
different weight of the background noise on the recorded signal. On
the other hand, this was probably one source of the overfitting found
in the initial, preliminary, model where the processed spike traces
were longer (7 ms) than those used in the final models (4 ms). Indeed,
in the presence of small and larger spikes with the same noise the
DL processing could have retained the background noise as a signature
of serotonergic neurons in addition to their shape and therefore this
may explain the improvement of the model obtained by shortening the
traces to be processed and limiting the recognition process to the
action potential shape. Notably, our synthetic model in which various background
masks were superimposed to 4 ms spikes did not significantly improve
the metrics compared to those of the biological model obtained using
original 4 ms spikes, confirming that limiting the DL process to the
spike was sufficient to eliminate the overfitting caused by the background
recognition together with the action potential shape for categorizing of neuron
type.
For this reason we considered sufficient to limit the model to recognition of short events and we did not include other parameters such as e.g. those defining the firing rhythm, in spite of the biological importance of this neuronal property. Indeed, our preliminary results (section 3.1) indicated that long segments of recording (e.g. 7 s), needed to allow the incorporation of periodicity of events in the model, resulted detrimental for the accuracy of the model itself. On the other hand, if deemed necessary for improving the accuracy and/or the complexity of the neuron classification, additional models directed to discriminate different, complementary, characteristics of each neuron class could be developed and then merged in a more refined model. 
For instance, a subset of putative serotonergic neurons recorded \emph{in vivo} displays complex firing in doublets or triplets (Hajos et al., 1995). Unfortunately, this specific activity is seldom observed in slices and our dataset of \emph{in vitro} recordings from genetically identified serotonergic neurons does not include any neuron with firing in doublets or triplets. Thus, our model may result not adequate to classify these neurons as serotonergic and should be modified to comply with this need. Nevertheless, it is likely that such neurons would not be missed even by our model developed to recognize single spikes because the interval between the two spikes in a doublet is usually greater than 3 ms (Hajos et al., 1995). Thus, the first spike will be fully comprised in the 4 ms detection window (of which $\sim$3 ms after upstroke) and recognized before the beginning of the second spike. In addition, when solitary spikes flank the doublets the recognition can be confirmed on these spikes. Thus, with minor fine tuning of parameters during training, the deep-learning procedure here described would be set to recognize also burst-firing neurons.
Altogether these considerations suggest that robust models based on CNN deep-learning procedures could be developed for specific application in conditions of recording where spikes of different amplitudes and possibly slightly variable shapes could be recorded from the same neuron as typical for in vivo recordings while the neuron is approached by a micropipette or in long duration recordings. The favourable characteristic of the model is that recognition of the neuron type can be performed at the beginning of the experiment on a limited number of spikes until the neuron is classified. Similarly, these models may be applied in high-density recordings in which special probes (e.g. silicon probes) allow simultaneous recording of hundreds of neurons in brain areas where different neuron types coexist. A model trained to recognize spikes from specific neurons would enable very rapid identification of the neurons captured in the different recording channels.

\paragraph{Perspectives}

Importantly, a relatively low number of recordings was sufficient
to develop our deep-learning based model. In perspective, the procedure
we describe can be applied to construct further models for the identification
of other spontaneously active monoaminergic neurons. For instance,
our approach with genetically fluorescent mice can be extended to
the recognition of other neurons for \emph{in vitro} recordings. Similarly,
application of the CNN deep-learning procedure to neuronal types recognized
with optogenetic methods (Liu et al., 2014) or with post-hoc immunohistochemistry
(Allers and Sharp, 2003) \emph{in vivo} may enable to construct a
template of models capable to recognize a variety of neurons during
\emph{in vivo} recordings from mouse and rats. Once validated, these
models would allow rapid identification of the recorded neuron, making
\emph{in vivo} recording of the activity of selected neurons more
feasible and less demanding than at present. This may also facilitate
studies on the correlation between the firing of different neuron
types and behavioural responses in laboratory animals and increase
our understanding of the physiological role of these neurons in modulating
higher brain functions.

In conclusion, our model provides the first proof of concept that
neurons can be recognized from the sole characteristics of extracellularly
recorded action potentials and independently of their firing rhythm. Our model
could readily be applied for intra-experiment decision making on the
experimental design to apply to record that specific neuron and/or
for helping the training of young Researchers at the beginning of
their experience.

\section{Acknowledgments}

The original recordings and measurements of the action potentials were performed
by Dr. Boris Mlinar and Dr. Alberto Montalbano. 

\section{References}
\begin{quote}
Allers, K.A., Sharp, T., 2003. \emph{Neurochemical and anatomical
identification of fast- and slow-firing neurones in the rat dorsal
raphe nucleus using juxtacellular labelling methods in vivo.} Neuroscience,
122(1),193-204. doi:10.1016/s0306-4522(03)00518-9.

Andrade, R., Haj-Dahmane, S., 2013. \emph{Serotonin neuron diversity in the dorsal raphe}. ACS
Chem Neurosci. 4(1), 22-5. doi: 10.1021/cn300224n.

Baraban, J.M. and Aghajanian, G.K., 1980. \emph{Suppression of firing
activity of 5-HT neurons in the dorsal raphe by alpha-adrenoceptor
antagonists}. Neuropharmacology 19, 355--363. doi:10.1016/0028-3908(80)
90187-2

Calizo, L. H.; Akanwa, A.; Ma, X.; Pan, Y.; Lemos, J. C.; Craige,
C.; Heemstra, L. A.; Beck, S. G., 2011. \emph{Raphe Serotonin Neurons
Are Not homogeneous: Electrophysiological, Morphological and Neurochemical
Evidence}. Neuropharmacology, 61 (3), 524\textminus 543. doi:10.1016/j.neuropharm.2011.04.008

Commons, K.G., 2020. \emph{Dorsal raphe organization}. J Chem Neuroanat.
110, 101868. doi: 10.1016/j.jchemneu.2020.101868.

De Luca, D., Moccia, S., Lupori, L., Mazziotti, R., Pizzorusso, T.,
Micera, S., 2023. \emph{Convolutional neural network classifies visual
stimuli from cortical response recorded with wide-field imaging in
mice}. J Neural Eng. 20(2), 026031. doi:10.1088/1741-2552/acc2e7.

Faulkner, P., Deakin, J.F., 2014.\emph{The role of serotonin in reward,
punishment and behavioural inhibition in humans: insights from studies
with acute tryptophan depletion}. Neurosci Biobehav Rev. 46 Pt 3:365-78.
doi:10.1016/j.neubiorev.2014.07.024. 

Gaspar P, Lillesaar C. \emph{Probing the diversity of serotonin neurons}.
Philos Trans R Soc Lond B Biol Sci. 2012 Sep 5;367(1601):2382-94.
doi:10.1098/rstb.2011.0378.

Hajós, M., Gartside, S.E., Villa, A.E., Sharp, T., 1995 \emph{Evidence for a repetitive (burst) firing pattern in a sub-population of 5-hydroxytryptamine neurons in the dorsal and median raphe nuclei of the rat}. Neuroscience 69(1):189-97. doi:10.1016/0306-4522(95)00227-a.

Johansen, A.R., Jin, J., Maszczyk, T., Dauwels, J., Cash, S.S., Westover,
M.B., 2016. \emph{Epileptiform Spike Detection Via Convolutional Neural
Networks}. Proc IEEE Int Conf Acoust Speech Signal Process. March
2016, 754-758. doi: 10.1109/ICASSP.2016.7471776.

Lesch, K.P., Araragi, N., Waider, J., van den Hove, D., Gutknecht,
L., 2012. \emph{Targeting brain serotonin synthesis: insights into
neurodevelopmental disorders with long-term outcomes related to negative
emotionality, aggression and antisocial behaviour}. Philos Trans R
Soc Lond B Biol Sci. 367(1601), 2426-43. doi:10.1098/rstb.2012.0039.

Levine, E.S., Jacobs, B.L., 1992. \emph{Neurochemical afferents controlling
the activity of serotonergic neurons in the dorsal raphe nucleus:
microiontophoretic studies in the awake cat}. J. Neurosci. 12, 4037--4044.
doi:10.1523/JNEUROSCI.12-10-04037.1992

Litjens, G., Kooi, T., Bejnordi, B.E., Setio, A.A.A., Ciompi, F.,
Ghafoorian, M., van der Laak, J.A.W.M., van Ginneken, B., Sánchez,
C.I., 2017.\emph{ A survey on deep learning in medical image analysis}.
Med Image Anal. 42, 60-88. https://doi.org/ 10.1016/j.media.2017.07.005. 

Liu, Z., Zhou, J., Li, Y., Hu, F., Lu, Y., Ma, M., Feng, Q., Zhang,
J.E., Wang, D., Zeng, J., Bao, J., Kim, J.Y., Chen, Z.F., El Mestikawy,
S., Luo, M., 2014. \emph{Dorsal raphe neurons signal reward through
5-HT and glutamate}. Neuron. 81(6),1360-1374. doi:10.1016/j.neuron.2014.02.010.

Liu, Y., 2018. \emph{Feature Extraction and Image Recognition with
Convolutional Neural Networks}. Journal of Physics: Conference Series,
1087. https://doi.org/10.1088/1742-6596/1087/6/062032.

Liu, L., Yang, S., \& Shi, D., 2019. \emph{Advanced Convolutional
Neural Network With Feedforward Inhibition}. 2019 International Conference
on Machine Learning and Cybernetics (ICMLC), 1-5. https://doi.org/10.1109/ICMLC48188.2019.8949229.

Mlinar, B., Montalbano, A., Piszczek, L., Gross, C., Corradetti, R.,
2016. \emph{Firing properties of genetically identified dorsal raphe
serotonergic neurons in brain slices}. Front. Cell Neurosci. 10, 195.
doi:10.3389/fncel.2016.00195 

Montalbano, A., Waider, J., Barbieri, M., Baytas, O., Lesch, K.P.,
Corradetti, R., Mlinar, B., 2015. \emph{Cellular resilience: 5-HT
neurons in Tph2 (-/-) mice retain normal firing behaviour despite
the lack of brain 5-HT}. Eur. Neuropsychopharmacol. 25, 2022--2035.
doi:10.1016/j.euroneuro.2015.08.021

Monti, J.M., 2011. \emph{Serotonin control of sleep-wake behavior.}
Sleep Med Rev. 15(4), 269-81. doi: 10.1016/j.smrv.2010.11.003. 

Mosko, S.S., Jacobs, B.L., 1974. \emph{Midbrain raphe neurons: spontaneous
activity and response to light}. Physiol. Behav. 13, 589--593. doi:10.1016/0031-9384(74)90292-3

Mosko, S.S., Jacobs, B.L., 1976. \emph{Recording of dorsal raphe unit
activity in vitro}. Neurosci. Lett. 2: 195--200.doi:10.1016/0304-3940(76)90014-8

Paquelet, G.E., Carrion, K., Lacefield, C.O,, Zhou, P., Hen, R., Miller,
B.R., 2022. \emph{Single-cell activity and network properties of dorsal
raphe nucleus serotonin neurons during emotionally salient behaviors}.
Neuron 110(16), 2664-2679.e8. doi: 10.1016/j.neuron.2022.05.015.

Pilowsky, P.M., 2014. \emph{Peptides, serotonin, and breathing: the
role of the raphe in the control of respiration}. Prog Brain Res.
209, 169-89. doi:10.1016/B978-0-444-63274-6.00009-6.

Rácz, M., Liber, C., Németh, E., Fiáth, R., Rokai, J., Harmati, I.,
Ulbert, I., Márton, G., 2020. \emph{Spike detection and sorting with
deep learning}. J Neural Eng. 17(1):016038. doi: 10.1088/1741-2552/ab4896.

Steinbusch, H.W.M., Dolatkhah, M.A., Hopkins, D.A., 2021. \emph{Anatomical
and neurochemical organization of the serotonergic system in the mammalian
brain and in particular the involvement of the dorsal raphe nucleus
in relation to neurological diseases}. Prog Brain Res. 261, 41-81.
doi: 10.1016/bs.pbr.2021.02.003. 

Tseng, H.A., Han, X., 2021. \emph{Distinct Spiking Patterns of Excitatory and Inhibitory Neurons and LFP Oscillations in Prefrontal Cortex During Sensory Discrimination}. Front. Physiol. 12:618307. doi: 10.3389/fphys.2021.618307.

Vandermaelen, C.P., Aghajanian, G.K., 1983. \emph{Electrophysiological
and pharmacological characterization of serotonergic dorsal raphe
neurons recorded extracellularly andintracellularly in rat brain slices}.
Brain Res. 289, 109--119. doi: 10.1016/0006-8993(83)90011-2.
\end{quote}

\section{Data Sharing Statement}

The data and source code correspondent to the analyses contained in
this manuscript are publicly available from: Corradetti et al. 2024
Repository with all data available at 

\texttt{GitHub.com/neuraldl/DLAtypicalSerotoninergicCells/tree/main} 

\section{Author Contributions }

All authors have made a significant contribution to the idea formation,
study design, data curation, analysis and interpretation. D.C. \&
R.C. wrote and reviewed the manuscript. R.C. collected and selected
the data. A.B. and D.C. realised the software for both the neural
networks and the synthetic data generation.
\end{document}